\documentstyle[prl,aps,floats,twocolumn]{revtex}
\input psfig.tex
\begin{document}
\twocolumn[\hsize\textwidth\columnwidth\hsize\csname @twocolumnfalse\endcsname
\title{A Causal Source which Mimics Inflation}
\author{Neil Turok}
\address{DAMTP, Silver St,\\
 Cambridge, CB3 9EW,  U.K.\\
Email: N.G.Turok@damtp.cam.ac.uk\\
}
\date{22/7/96}
\maketitle

\begin{abstract}
How unique are the inflationary predictions for the 
cosmic microwave anisotropy pattern? In this paper, 
it is asked whether an arbitrary 
causal source for perturbations in the standard hot big bang
could effectively mimic the predictions of the 
simplest inflationary models. A surprisingly simple
example of a `scaling' causal source is found to
closely reproduce the inflationary predictions. 
This Letter extends the work of a previous paper (ref. 6) to
a full computation of the anisotropy pattern, including
the Sachs Wolfe integral. I speculate on the possible physics
behind such a source.
\end{abstract}
\hspace{0.2in}
]

The prospect of mapping the cosmic microwave background (CMB) 
anisotropies to high resolution
has raised the exciting possibility of confirming fundamental 
theories of the origin of structure in the universe. 
The inflationary theory, being the simplest and most complete,
is the present front-runner, and the latest CMB
measurements do even seem to hint at the presence of 
first Doppler peak in the angular power spectrum, at 
$l\sim 200$ as 
predicted  by the simplest, spatially flat
inflationary models.
These spectra are distinct
from those predicted by cosmic defect or baryon isocurvature models,
and it is
an important question whether 
spectra of this form are 
really
a unique prediction of inflation. 
Or could a non-inflationary 
mechanism somehow replicate them?

The fundamental difference between  inflationary and
non-inflationary mechanisms of structure formation is that
inflation alters 
the causal structure of the early universe, adding on a
prior epoch during which correlations are established on scales
much larger than the Hubble radius. 
This is of course how the standard `horizon puzzle' is solved.
The perturbations produced during inflation are susceptible to very 
detailed test, and one could hope that 
the same feature of 
`super-horizon' scale correlations (quotes indicate a standard big
bang definition) could be used as a signature.
If `super-horizon' perturbations were shown to exist,
this would strongly
confirm
inflationary structure formation,
for no causal mechanism could have produced them
within the standard big bang.

Since COBE observed 
large scale perturbations on the CMB sky,
one might think this already 
showed that `super-horizon' perturbations 
were present at last scattering. 
But these large angle 
anisotropies could have been
produced
causally within the standard big bang, through a time dependent 
gravitational potential along the line of sight.
Cosmic defects, as well as  open universe
or $\Lambda$ dominated models
provide explicit examples
of theories
where this actually happens.

\begin{figure}[htbp]
\centerline{\psfig{file=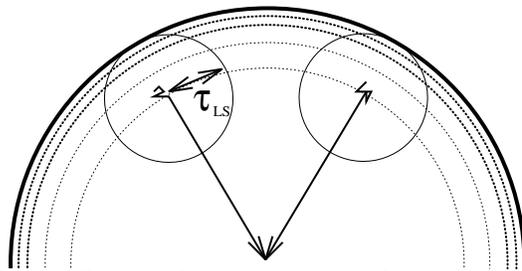,width=2.7in}}
\caption{ The causality constraint on the microwave background anisotropy. 
The picture is in comoving coordinates:
the outer circle represents our causal horizon in the standard big bang.
The inner circle is the surface of last scattering, on which
photons are `set free' from the hot plasma on their path to us. 
Circles show the domains 
of influence on these photons - the radius
is the light travel distance 
since the hot big bang $\tau_{LS}$. This subtends an angle $\Theta_{LS}
\sim 1.1^o$ in a flat universe with standard recombination.
No causal physics operating within the standard big bang
could have generated correlations between 
photons on the last scattering surface at
points separated by more than
$2 \Theta_{LS}$ on the sky.
}
\label{fig:f1}
\end{figure}

The smaller angle anisotropies are a more  promising probe
because they are due to local effects 
on the surface of last scattering, 
which are  strongly constrained by causality (Figure 1). 
In particular, the pattern of Doppler peaks
caused by phase coherent oscillations \cite{Sakharov} in the photon-baryon 
fluid provides a possible signature of `super-horizon' 
curvature perturbations.
Early papers exploring this include refs.
\cite{ct},
\cite{albrecht} and 
\cite{hw}. This Letter is a follow-up of \cite{ntfirst}, 
where an inflation-mimicking model  was proposed. 
A recent paper
by Hu, Spergel and White \cite{hsw} criticising that 
model is now undergoing revision.

In this Letter I ask whether a causal source acting 
purely via gravity within the standard hot big bang could
generate CMB
anisotropies similar to those in flat inflationary 
models. 
I further restrict consideration to `scaling' sources, 
mainly for simplicity.
This restricted set 
of models turns out to provide a surprisingly simple
inflationary mimic. I emphasise that the mimic is not a theory,
but simply an
`ansatz' constructed by hand to provide a
consistent solution to the Einstein equations. As such it provides
a counterexample to some arguments in the literature 
(see e.g. \cite{hw}). Furthermore it
has a sufficiently simple form that it  is not out of the question 
that it
might actually be realised in a future theory of structure
formation - in that sense the counterexample may turn out to 
be constructive.

I deal with the linearised Einstein equations in the `stiff'
approximation, in which the induced perturbations are assumed to
have negligible effect on the source \cite{VS}.
In this approximation, the source stress energy tensor $\Theta_{\mu \nu}$
is covariantly conserved with  respect to
the background metric:
\begin{equation}
\dot{\Theta}_{00} + \frac{\dot{a}}{a}(\Theta_{00}+\Theta) = \Pi ;
\quad
\dot{\Pi} + 2 \frac{\dot{a}}{a} \Pi = \partial_i\partial_j 
\Theta_{ij}
\label{eq:momc}
\end{equation}
where $\Pi = \partial_i \Theta_{0i}$. 
Dots denote derivatives with respect to conformal
time $\tau$, and $a(\tau)$ is the scale factor.

A formalism for dealing with such sources was proposed in
\cite{ntfirst}. Here I shall consider only `coherent' sources, 
which are representable in terms of 
a single set of `master functions'. 
These are ordinary functions, which obey 
(\ref{eq:momc}). 
The unequal time correlator $<\Theta_{\mu \nu}(r,\tau)
\Theta_{\rho \lambda}(0,\tau')>$ equals 
the spatial 
convolution of the master function for  $\Theta_{\mu \nu}$ 
with that for $\Theta_{\rho \lambda}$.

As argued in \cite{ntfirst}, 
the causality constraint implies that
the master functions
$\Theta_{\mu \nu}(r,\tau)$ are zero for all $r>\tau$. 
If we ignore vector and tensor perturbations, they
are spherically symmetric, and we can
represent them as 
$\Theta_{00}(r,\tau)$, $\Theta_{0i}= x_i J(r,\tau)$, 
$\Theta_{ij}= {1\over3} \Theta(r,\tau)\delta_{ij}+
\Theta^A(r,\tau)(x_ix_j-
{1\over 3} r^2 \delta_{ij}) $. The term ${1 \over 3} \Theta$ is 
the pressure $P$, and $\Theta^A$ the anisotropic stress. 
In situations where matter is being actively  moved,
as it will be here, the anisotropic stresses are generally
of the same order as the pressure.

Some general properties can now be seen. The Fourier transforms 
(assumed to exist) are analytic about 
$k_i=0$, and 
can be expanded in a Taylor series. The leading terms are then 
(by isotropy) 
$\Theta_{00} \sim k^0$, $\Theta_{0i} \sim k_i$ (from which 
$\Pi \sim k^2$ follows) and
$\Theta_{ij} \sim \delta_{ij}$. In Fourier space we write
$\Theta_{ij}({\bf k}) \equiv
{1\over 3} \delta_{ij} \Theta +(\hat{k}_i\hat{k}_j -{1\over 3} \delta_{ij})
\Theta^S$, where $\Theta^S = k^{-1} d\Theta^{A}/dk - d^2 \Theta^{A}/dk^2$.
It follows that 
$\Theta^S \sim k^2$.

I now specialise to `scaling' sources, in which 
one assumes that the source a) involves a number with 
dimensions of the inverse of Newton's constant $G$ 
and b) involves no other length scale apart from
the horizon scale $\tau$. If these conditions are met, the 
source-perturbation equations are scale-invariant,
apart from the violation of scaling caused by the 
radiation-matter transition.
Scaling and dimensional analysis 
imply that (see e.g. \cite{pst})
$\Theta_{00}(k,\tau) \sim \tau^{-{1\over 2}} f_1(k\tau)$,
$\Theta(k, \tau) \sim \tau^{-{1\over 2} } f_2(k\tau)$,
$\Pi(k, \tau) \sim \tau^{-{3\over 2} } f_3(k\tau)$,
$\Theta^S(k, \tau) \sim \tau^{-{1\over 2} } f_4(k\tau)$,
where the $f_i$ have power series
expansions in $k^2$ obeying the restrictions noted above. 
These four  $f_i's$ are are related by
the two energy momentum conservation equations  (\ref{eq:momc}).
So for example, the $k^0$ term in the first equation relates the 
leading
terms in $f_1$ and $f_2$, and the $k^2$ term in the 
second equation relates the leading terms in  $f_2$ and $f_3$.
But even after applying these equations, 
we still have essentially two free functions remaining. 
We also have some freedom in how
to incorporate the
matter-radiation transition into the source.

I assume the background spacetime is flat, 
and has metric $ds^2= a^2(\tau)(-d\tau^2+ 
(\delta_{ij}+h_{ij}(x,\tau))dx^idx^j)$,
with $\tau$ conformal time and
$a(\tau)$ the scale factor. I work in initially unperturbed
synchronous gauge, which is especially
suitable for causal theories since the Einstein 
equations are manifestly causal (in contrast, in Newtonian gauge 
the influence of anisotropic stresses propagates acausally \cite{VS}).
This gauge is also  
straightforward to interpret - for example the radiation density
contrast $\delta_R$ is just that measured 
in the rest frame of freely falling
particles like cold dark matter
particles.

We are interested in computing the CMB temperature distortion on the 
sky in a direction ${\bf n}$: 
\begin{equation}
\frac{\delta T}
{T}({\bf{n}})
 = \frac{1}{4}\delta_{R}(i) - {\bf n \cdot v_R}(i)
 - \frac{1}{2}\int_i^f d\tau \dot{h}_{ij} {\bf n}^i{\bf n}^j
\label{eq:s1}
\end{equation}
where $\delta_R$ is the
density contrast, and ${\bf v_R}$ the velocity, of the
photon fluid on 
the surface of last scattering.
The last term is the Sachs-Wolfe integral, 
representing
the change in the proper path length
field along the line
of sight. This expression holds 
in the `instantaneous recombination' approximation, which I shall use 
throughout. 

The first two terms are local effects on the surface of 
last scattering. 
They are determined from: 
\begin{equation}
\ddot{\delta_C} + {\dot{a}\over a} \dot{\delta_C} =
 4 \pi G\bigl(
\sum_N (1+3 c_N^2) \rho_N \delta_N + \Theta_{00}+\Theta \bigr),
\label{eq:c1}
\end{equation}
\begin{equation}
\dot{\delta_R} = {4\over 3}\dot{\delta_C} -{4\over 3} {\bf \nabla \cdot v_R}; 
\dot{\bf v_R} = - (1-3 c_S^2){\dot{a} \over a} {\bf v_R}
-{3\over 4}  c_S^2 {\bf \nabla } \delta_R
\label{eq:c2}
\end{equation}
where $c_S$ is the speed of sound in the photon-baryon fluid.
The only component of the stress energy tensor that enters
is $\Theta_{00}+\Theta$, i.e. $\rho+3P$.
Thus prior to last scattering, only
one of the two free functions in $\Theta_{\mu \nu}$ 
contributes - the other is literally `invisible' in the CMB 
anisotropy.

In the simplest inflationary theory, the surface 
term $ \frac{1}{4}\delta_R$ 
captures most of the relevant 
physics determining the location of the 
Doppler peaks. In ref. \cite{ntfirst}, 
I focussed only on this term, and found a causal source $\Theta_{00}+\Theta$
which 
mimicked the same term in the inflationary theory.
The idea was simply to choose 
\begin{equation}
\Theta_{00}+\Theta \propto  f_1(r)+f_2(r)
 \propto \delta(r-A\tau)
\qquad 0<A \leq 1
\label{eq:fun}
\end{equation}
representing a spherical shell expanding at some fraction of
the speed of light. I found that for $A$ close to unity,
the
$\delta_{R}$ surface term closely matched the inflationary one.
Such a shell of matter is similar in form to a supernova explosion -
for a spherical shell of 
neutrinos, one has $\Theta_{ij} \sim \Sigma p^i p^j 
\propto x^i x^j/r^2$. In that case,
$\Theta^S$ and
$\Theta$ are comparable in magnitude, and the same is true here.

Here I  
extend the computation 
to the entire expression 
(\ref{eq:s1}).
The Sachs Wolfe integral introduces some dependance
on the 
anisotropic part of the metric perturbation.
To compute this 
it is necessary to further specify $\Theta_{\mu \nu}$, by for example 
sepcifying another of 
the $f_i$ functions. The simplest choice leaving $f_1+f_2$ fixed 
is to specify $f_3$. Then equations (\ref{eq:momc}) are 
used as follows: the energy equation is integrated to determine
$\Theta_{00}$, and the momentum equation is differentiated to determine
$\Theta^S$. Of course this must be done consistently with 
the matching of the leading terms as discussed above. 

In Fourier space, the choices I make for the source  are:
\begin{equation}
\Theta_{00}+\Theta= {a\over \dot{a}} {{\rm sin} A k \tau 
\over A k \tau^{5\over 2}}
\label{eq:ans}
\end{equation}
as in (\ref{eq:fun}), with the prefactor 
incorporating the radiation-matter transition in a simple way.
For $\Pi$, we must satisfy 
$\Pi(k) \sim k^2$ at small $k$. Equivalently, 
the integral 
$\int_0^\tau r^2 dr \Pi(r,\tau)=0$. This is most easily satisfied by
taking $\Pi(r,\tau)$ to be the sum of two delta functions, of equal weight but
opposite sign. Their Fourier transform produces:
\begin{equation}
\Pi= - { E(\tau)\over k \tau^{5\over 2}} 
 {6 \over B^2-C^2} \bigl(
{ {\rm sin} B k \tau \over B} - { {\rm sin} C k \tau \over C}\bigr)
\label{eq:anp}
\end{equation}
where $E(\tau)$ is a messy function  
obtained by analytically 
solving for the coefficient of 
$k^2$ term in the momentum equation (\ref{eq:momc}).
It equals ${2\over 15}$ in the radiation era and ${2\over 18}$ in the matter
era. 
A set of values which  I find
leaves the Sachs Wolfe integral sub-dominant is $B=1.0$ and $C=0.5$. 

The anisotropic 
metric perturbation $h^S(k)$ (defined analogously to $\Theta^S$) is
now given by 
$\dot{h}^S-\dot{h}= -24 \pi G(\Pi+ \Sigma_N(P_N+\rho_N)a^2 i{\bf k}\cdot
{\bf v_N})/k^2 $. Finally, I model the free streaming 
of photons and neutrinos after last scattering following ref. \cite{ct}.
The initial conditions for the $\Theta_{00}$ and the perturbations 
are set up deep in the radiation era $\tau<<\tau_{EQ}$, and 
well outside the horizon $k\tau <<1$: they are read off from the
$k^0$ terms in the energy conservation, and perturbation 
equations: 
\begin{equation}
\Theta_{00} = 2 \tau^{-{1\over 2}} \qquad \delta_R = \delta_\nu= {4\over 3}
\delta_C = D \tau^{3\over 2}  \qquad  {\bf v}_R =0 
\label{eq:inc}
\end{equation}
with the constant $D$ determined by setting the total pseudoenergy 
$\tau_{00}= k^2(h-h^S)/(24 \pi G)=\Theta_{00} + \sum_N \rho_N a^2
 \delta_N +(\dot a / a) \dot{\delta_C} /
(4 \pi G)$ to zero. With  these choices there
are no superhorizon perturbations in the 
photon-to-CDM, 
photon-to-baryon or photon-to-neutrino ratios. The
pseudoenergy is just
the Ricci scalar of the spatial slices, so setting it zero means
there are no curvature perturbations either. 
These initial conditions 
are thus {\it both} `adiabatic' and `isocurvature' - there
are simply {\it no} perturbations in the universe on superhorizon scales.
The complete $C_l$ spectrum of the causal model defined 
in equations (6-8) is shown
in Figure 2.

\begin{figure}[htbp]
\centerline{\psfig{file=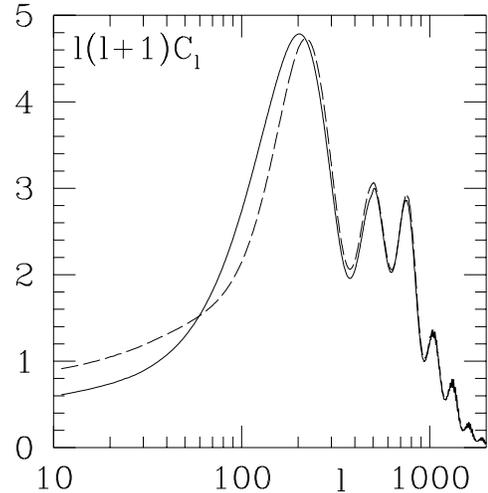,width=2.7in}}
\caption{ Comparison of the simplest inflationary theory 
with its `mimic' causal source model 
discussed here. The vertical axis is $l(l+1) C_l$,
with $C_l$ the angular power spectrum, and $l$ 
the Legendre index. 
Both curves were calculated in the same
`instantaneous recombination' approximation, in a flat universe
with
`canonical' parameters $\Omega_B=0.05$, $\Omega_{CDM}=0.95$.
The vertical scale is arbitrary.
}
\label{fig:f2}
\end{figure}

What about the sub-horizon behaviour of the source? 
In the construction above, where I integrate the energy equation to
determine $\Theta_{00}$, there is no reason it should tend to zero
well inside the horizon. However, because I explicitly turn off
$\Theta_{00}+\Theta$ inside the horizon, 
the source ceases to have any effect on the fluid perturbations and
the trace part of
the metric $h = -2 \delta_C$. The effect on the 
anisotropic part $h^S$ is similarly turned off 
because $\Pi$ goes to zero. In effect, I have turned off
all the `gravitationally active' components of the source,
but there is no reason for the energy  $\Theta_{00}$, the
pressure  $\Theta$ or the anisotropic stresses  $\Theta^S$
to separately vanish - they only need to satisfy the relations 
$\Theta_{00} +\Theta = 0$ and $\Theta + 2 \Theta^S=0$. 
This is reminiscent of the behaviour of a straight
cosmic string - it carries energy but generates 
no gravitational field. In any case, the sub-horizon source is removed
if one adds 
a term  
$c_1 k^2 \tau (a/\dot{a}) \Theta_{00}$ 
to $\Theta_{00} +\Theta$, or a term $- c_2 k^2 \tau \Theta_{00}$ to $\Pi$.
Either of these makes
$\Theta_{00}$, $\Theta$, and  $\Theta^S$ go to zero inside the 
horizon as exp($-const. k^2 \tau^2$). 
Figure 3 shows the evolution of the components of $\Theta_{\mu \nu}$ 
with and without these modifications, 
and Figure 4 shows the corresponding $C_l$ spectra.
The  moral is that there is a lot of freedom
inside the horizon to make very large 
alterations in the source without significantly 
affecting the CMB anisotropies.

\begin{figure}[htbp]
\centerline{\psfig{file=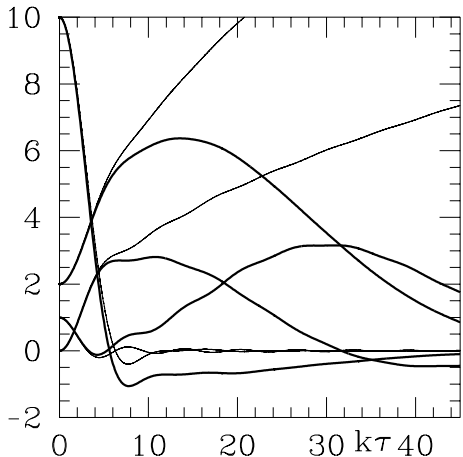,width=2.5in}}
\caption{ Behaviour of the stress tensor
inside the horizon. The thin lines show the original source
defined in equations (6) and (7), as a function 
of $k\tau$ in the radiation era 
(in order as the curves 
intersect the y axis moving upwards) $\tau^{1\over 2}\Theta_S$,
$\tau^{1\over 2}(\Theta+\Theta_{00})$, 
$\tau^{1\over 2}\Theta_{00}$ and $75 k^{-2}\tau^{-{1\over 2}}
\Pi$. The bold lines show the same curves for the 
model specified by $c_1=.001$ and $c_2=.003$ ( see text),
where all components are `turned off' inside the horizon.
The $C_l$ spectrum for the latter model is shown 
in Figure 4.
}
\label{fig:f3}
\end{figure}

How sensitive to the particular choice of Ansatz is the result?
Figure 4 compares the $A=1$ model with the cases $A=0.7$,
and $A=0.1$. The leftward shift in the first peak was 
explained in ref. \cite{ntfirst}. As shown there, 
it is also easy to arrange for
a shift to the right. So while the $A=1$ model is on the
causality limit, there is a substantial region of parameter space around it with
similar $C_l$ spectra.
It would
be very difficult
to observationally separate these models
from inflation, especially in the realistic
case where we are unlikely to know all the relevant 
cosmological parameters
($\Omega$, $\Omega_B$, $h$, $\Omega_\Lambda$
and so on) in advance. 
I have so far explored only a very tiny part of model space,
and preliminary investigations of more general Ansatzes 
indicate that a very wide range of $C_l$ spectra are possible,
especially when the Sachs Wolfe integral becomes a dominant effect.

\begin{figure}[htbp]
\centerline{\psfig{file=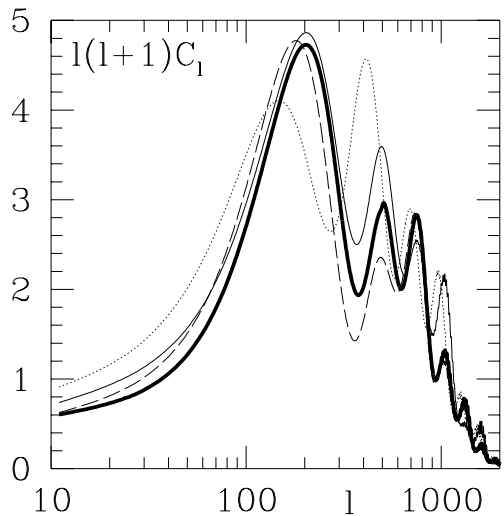,width=2.8in}}
\caption{ The $C_l$ spectrum of the 
$A=1$ model (bold line) is compared with that for $A=0.7$
(dashed line) and $A=0.1$ (dotted line). The model illustrated in
Figure 3, for which the stress tensor vanishes inside the horizon,
is also shown as the thin line (altering $c_2$ to $0.001$ makes the
$C_l$ spectrum virtually indistinguishable from the original model,
 except for
the fourth peak, which is still a little  high)).
}
\label{fig:f4}
\end{figure}

I have given myself a great deal of freedom in
constructing this source, and it is far from clear that
realistic physics could produce it. 
Nevertheless, the reader may tolerate some speculations.
As mentioned, the form of $\rho+3P$ required 
is similar to that resulting from a supernova explosion, but
whereas in the latter case the energy redshifts away as $a(t)^{-1}$,
here, scaling evolution requires that the energy in the shell
increases - by dimensions, $E \sim G^{-1} t$.
This requires some 
positive feedback mechanism, which one could conceivably arrange
with unstable dark matter, decaying via stimulated emission
of Goldstone bosons, or even gravity waves.
Another issue is the Gaussianity or otherwise
of the perturbations. If the source is
made up of a very large number of `explosions' which are 
allowed to superpose, then it can be made arbitrarily Gaussian. 
But if the exploding shells interact, there would be a limit to their
number density, and the perturbations would then be 
nonGaussian.

The conclusion of this paper is that 
causality alone is insufficient to distinguish the inflationary
$C_l$ predictions from those of non-inflationary models.
Of course the observational confirmation of one of these spectra 
would
be a tremendous success for inflation, but 
the door would nevertheless still be left open to other 
possible 
explanations of cosmic structure formation.

I thank A. Albrecht and J. Maguiejo for discussions, 
R. Crittenden for collaboration on the codes
used here, 
and W. Hu, D. Spergel and M. White for helpful correspondence.
This work was supported by a grant from Cambridge University and 
PPARC, UK.

\end{document}